\documentclass[12pt]{article} 

\usepackage{mathbbol,amssymb,latexsym,amsfonts,amsmath,amsthm}

\usepackage{amsfonts}
 \usepackage{amssymb}
 \usepackage{amsmath}

\def\pmx{\begin{pmatrix}}
\def\emx{\end{pmatrix}}
\def\bsq{\begin{subequations}}
\def\esq{\end{subequations}}
\def\be{\begin{eqnarray}}
\def\ee{\end{eqnarray}}
\def\bee{\begin{eqnarray*}}
\def\eee{\end{eqnarray*}}
\def\bal{\begin{align}}
\def\eal{\end{align}}

\newtheorem{thm}{Theorem}
\newtheorem{lemma}[thm]{Lemma}
\newtheorem{prop}[thm]{Proposition}
\newtheorem{conj}[thm]{Conjecture}

\def\bra{\langle}
\def\ket{\rangle}

\def\kb{ \ket \bra }

\def\half{{\textstyle \frac{1}{2}}}

 \def\tr{\hbox{Tr} \,}
 
\def\mm{ \! - \! }

\def\nn{\nonumber}
\def\ot{\otimes}

\def\wtd{\widetilde}

\def\ovb{\overline}

\def\s2{\tfrac{s}{2}}

\def\mm{ \! - \! }

\def\qed{\qquad{\bf QED}}
\def\pf{ \noindent{\bf Proof:} }

\newcommand{\proj}[1]{ | #1 \kb  #1|}

\title{\bf \large Connecting $N$-representability to  Weyl'sÊ problem: \\
The one particle density matrix for $N = 3$ and $R = 6$ }

\author{ Mary Beth Ruskai  \thanks{Partially supported   
   by the National Science  Foundation under Grant  DMS-0604900} 
  \\  {\small Department of Mathematics,
     Tufts University,
       Medford, MA 02155} \\
     {\small     Marybeth.Ruskai@tufts.edu}}
     
\begin{document}
 
 \maketitle

 \begin{abstract} 
An analytic proof is given of the necessity of the Borland-Dennis
conditions for $3$-representability of a one particle density
matrix with rank 6.    This may shed some light on Klyachko's
 recent use of  Schubert calculus to find general conditions
 for $N$-representability.
 \end{abstract}
 
 \section{Introduction}
 
The recent announcement by A. Klyachko \cite{K1} of the solution
 of the pure state $N$-representability problem for fermionic one-particle
 density matrix observes that this is the first new result since the work of
 Borland and Dennis \cite{BD} in the early 1970's.    There may therefore
 be some historical value in unpublished work of the author from that time, which
 makes a connection between the Borland-Dennis conditions and Weyl's
 problem.    The latter asks for conditions on sequences $\{a_k\}, \{b_k \}, \{c_k\} $
 which ensure that there exist self-adjoint matrices
 $A, B, C$  with eigenvalues $a_k, b_k, c_k$ respectively such that $A+B = C$.    
The first complete solution to Weyl's problem was given by Klyachko \cite{K2}
 in  1998.
 
 Let $\gamma$ be a density matrix normalized so that $\tr \gamma = N$.
 The pure state  \linebreak $N$-representability problem for fermions asks for necessary
 and sufficient conditions on $\gamma$ for the existence of an antisymmetric $N$-particle
 state whose one-particle reduced density matrix is $\gamma$.    Let  $R$ 
 denote the rank of $\gamma$.   For the case $N = 3$ and $R = 6$, Borland and
 Dennis gave a  pair of conditions on the eigenvalues $\lambda_k$ of $\gamma$
 which can be written as follows under the assumption that they 
 are arranged in non-increasing order.
 \be   \label{BDeq}
     \lambda_1 + \lambda_6 = 1, \quad  \lambda_2 + \lambda_5= 1,  \quad  \lambda_3 + \lambda_4 = 1
     \ee
     \be
      \lambda_1 + \lambda_2 \leq \lambda_3 + 1   \label{BDineq}
 \ee
Note that \eqref{BDeq} can be written compactly as $   \lambda_k + \lambda_{7-k} = 1$
for $ k = 1,2,3 $

Borland and Dennis \cite{BD} proposed their conditions on the basis of
numerical studies and gave a proof of \eqref{BDineq} under an assumption,
which is equivalent to \eqref{BDeq}, about the pre-image of $\gamma$.
In this note, we show that \eqref{BDeq} is a necessary condition for  \linebreak
$N$-representability
when $N = 3$  and $R = 6$, completing the analytic proof of  Borland and Dennis.
We begin with some  background material in Section~\ref{sect:back}.
In Section~\ref{sect:cond1} we present a proof of the necessity of \eqref{BDeq}.
In Section~\ref{sect:cond2} we give a different, independent proof of the necessity 
of the inequality \eqref{BDineq} from Weyl's inequalities.   For completeness,
we include a proof of sufficientcy of \eqref{BDeq} and \eqref{BDineq} in
Section~\ref{sect:suff}.     In Sections~\ref{sect:R=N+3}  and \ref{sect:Weyl} we 
present some partial results for the cases $N = 3$ and $ R = N + 3$ in the hope of providing
some intuition behind the success of Klyachko's approach to a full solution.

\section{Notation and background} \label{sect:back}

In this note, we write the eigenvectors of $\gamma$ as $|\phi_k \ket$
so that
\be   \label{dm}
    \gamma = \sum_k \lambda_k \proj{\phi_k} .
\ee
We will  let ${\cal A} $ denote the anti-symmetrization operator and
use the notation  $[f_j,f_k  , f_{\ell}] = {\cal A}  f_j(x_1) f_k (x_2)  f_{\ell}(x_3) $
 to denote a  Slater determinant.  
The notation $\bra ~ , ~ \ket_m$  indicates a partial inner product on a tensor 
product of Hilbert spaces.

We need some results from Section 10 of Coleman's fundamental paper \cite{Cole}.   
The first is Theorem~10.6 in \cite{Cole}
\begin{lemma} {\em (Coleman)}  \label{thm:cole}
The one-particle density matrix $\gamma$ is $N$-representable with pre-image
$|\Psi \ket =    \sqrt \lambda_1  \, {\cal A} \, |\phi_1   \ket  \ot  |\Phi_1 \ket  + 
   \sqrt{1 - \lambda_1} \, |\Phi_2 \ket $
 if and only if it can be written in the form 
\be   \label{cole}
    \gamma = \lambda_1 \proj{\phi_1} +  \lambda_1 \gamma_1 + (1 -  \lambda_1) \gamma_2
\ee
where $\gamma_1$ is the $(N \mm 1)$-representable reduced density matrix of
$|\Phi_1 \ket$, and $ \gamma_2$ is $N$-representable with  pre-image $\Phi_2$
satisfying
\be   \label{stong}
    \bra \phi_1, \Phi_2 \ket_1 =  \bra \Phi_1, \Phi_2 \ket_{2,3 \ldots N} = 0.
\ee
\end{lemma}
The next two results are Theorems~10.2 and 10.4 respectively in \cite{Cole}.
  (See also  \cite{RuskPH}.)
\begin{thm}  \label{N=2}
A one-particle density matrix $\gamma$ is $2$-representable
 if and only if all non-zero eigenvalues are doubly degenerate.
 If there are no other degeneracies and the eigenvalues are 
 written in non-increasing order so that 
 $\lambda_{2k-1} = \lambda_{2k} > \lambda_{2k+1} $,
 then the pre-image of $\gamma$ must have the form
 \be
  |  \Psi \ket = \sum_k e^{i \theta_k} \sqrt{\lambda_{2k}} \,  [\phi_{2k-1}, \phi_{2k}]
 \ee
\end{thm}
\begin{thm} \label{thm:ph} When $N = 2n+1$ is odd and the one-particle density matrix 
$\gamma$ has rank  $R = N + 2$, it is
 $N$-representable if and only if $\lambda_1 = 1$  and the
remaining eigenvalues are doubly degenerate.     
\end{thm}

\section{Necessity of the condition   $   \lambda_k + \lambda_{7-k} = 1$.}   \label{sect:cond1}

To show that \eqref{BDeq} is a necessary condition for 3-representability when
$R = 6$, observe that since $\lambda_1  = \bra \phi_1,  \gamma \phi_1 \ket$
it follows from \eqref{cole}  that
\bee
   \bra  \phi_1,  \gamma_1 \, \phi_1 \ket =   \bra  \phi_1,  \gamma_2 \, \phi_1 \ket  = 0.
\eee
Therefore, $\gamma_1$ and $\gamma_2$ have rank $\leq 5$.   It then follows from
Theorem~\ref{thm:ph} that one can write
\bee
     \gamma_2 = \proj{g_1} +  |a|^2  \proj{g_2} +   |a|^2  \proj{g_3}  +  |b|^2  \proj{g_4} 
         +  |b|^2  \proj{g_5} 
\eee
with $|a|^2 + |b|^2 = 1$ and  $    |\Phi_2\ket = a [g_1,g_2,g_3] + b [g_1,g_4,g_5] $.
There is no loss of generality in writing  $\Phi_1 =   \sum_{j < k} c_{jk} [g_j,g_k]$.

We first consider the case in which both $a,b \neq 0$.
Then a simple computation shows that  \eqref{stong} implies
\bee
   | \Phi_1 \ket = c_{24}  [ g_2,g_4] + c_{25}  [ g_2,g_5] + c_{34}  [ g_3,g_4] + c_{35}  [ g_3,g_5] 
   \eee
  so that $\bra g_1, \Phi_1 \ket_1 = 0$.   Defining $|\phi_6 \ket = |g_1\ket$, gives
  $\lambda_6 = 1 - \lambda_1$ and
 one can rewrite 
   \eqref{cole} as
\be    \label{colealt}
    \gamma = \lambda_1 \proj{\phi_1} + (1- \lambda_1) \proj{\phi_6} +
             \lambda_1 \gamma_1 + (1 -  \lambda_1) \wtd{\gamma}_2
\ee
where $\wtd{\gamma}_2 = \gamma_2 - \proj{g_1}$ is the reduced density matrix
of $|G_1 \ket = \bra g_1, \Phi_2 \ket_3 = a  [ g_2,g_3] + b [ g_4,g_5] $.   
Thus, in the orthonormal basis $\{ g_2, g_3 , g_4, g_5 \}$ we find
\bee
     \gamma_1 = \pmx  |c_{24} |^2 + |c_{25} |^2  &  \ovb{c}_{24} c_{34} +   \ovb{c}_{25} c_{35} 
         & 0 & 0 \\  c_{24} \ovb{c}_{34} +   c_{25} \ovb{c}_{35}  &  |c_{34} |^2 + |c_{35} |^2 & 0 & 0 \\
         0 & 0   &  |c_{24} |^2 + |c_{34} |^2 &  \ovb{c}_{24} c_{25} +   \ovb{c}_{34} c_{35}  \\
          0 & 0   &   c_{24} \ovb{c}_{25} +  c_{34} \ovb{c}_{35}     &  |c_{25} |^2 + |c_{35} |^2  \emx .
  \eee
The key point is that $\gamma_1$ is block diagonal
and can be diagonalized by a block diagonal unitary transformation
which mixes only within pairs $g_2, g_3$ and $g_4,g_5$ leaving the Slater
determinants in $G_1$ unaffected (except possibly for a phase factor which
can be absorbed into the new basis).
Denoting the new basis as $ \phi_k$, 
we now have $|G_1 \ket = a [  \phi_2, \phi_3] + b [  \phi_4, \phi_5] $.
 Then either by explicit computation
or from Coleman's proof \cite{Cole2} of Theorem~\ref{N=2}, one can write
$|\Phi_1 \ket = s  [  \phi_2, \phi_4] + t [  \phi_3, \phi_5] $ with $|s|^2 + |t|^2 = 1$.
Thus, the eigenvalues of $\gamma$ satisfy
\bsq  \label{eval24} \be
    \lambda_2 & = & \lambda_1 |a|^2 + (1- \lambda_1) |s|^2   \\
     \lambda_3 & = & \lambda_1 |b|^2 + (1- \lambda_1) |s|^2 \\   
    \lambda_4 & = & \lambda_1 |a|^2 + (1- \lambda_1) |t|^2  \\
    \lambda_5 & = & \lambda_1 |b|^2 + (1- \lambda_1) |t|^2 
\ee  \esq 
which implies
\be  \label{sum}
      \lambda_2  +   \lambda_5 =   \lambda_3 +   \lambda_4 =   \lambda_1 + (1 -   \lambda_1 ) = 1.
\ee

We now consider the possibility that one of $a, b$ is zero, in which case,
$| \Phi_2 \ket $ is a single Slater determinant and there is no loss of generality
in writing as $\Phi_2 = [g_1,g_2,g_3]$.   Then \eqref{stong} implies that
one can write
\be
     | \Phi_1 \ket = \sum_{j =1,2,3} \sum_{k = 4,5}  x_{jk}  [g_j,g_k] +  c [g_4,g_5]
\ee
Now regard $x_{jk}$ as a $3 \times 2$ matrix and observe when $U,V$ are
$3 \times 3$ and $2 \times 2$ unitary matrices, $Y =ÊU X V^\dag$ 
corresponds to a basis change
which mixes  $g_1, g_2, g_3$ and $g_4, g_5$ among themselves.   By
the singular value decomposition we can find $U, V$ such that only
$y_{24}$ and $y_{35}$ are non-zero.   Thus, in the new
basis  which we call $\phi_k$ 
\be
   | \Phi_1 \ket  = y_{24} [\phi_2,\phi_4] + y_{35} [\phi_3,\phi_5] + c  [\phi_4,\phi_5].
\ee
Again writing $\phi_6 = g_1$, we find that the pre-image of $\gamma$ has the form
\be
  |\Psi \ket = a_{123} [\phi_1,\phi_2,\phi_3] + a_{246}[\phi_2,\phi_4,\phi_6] +
      a_{356}[\phi_3,\phi_5,\phi_6]  +  a_{456}[\phi_4,\phi_5,\phi_6] 
\ee
which implies \eqref{BDeq}.

\section{Necessity of the inequality \eqref{BDineq}}  \label{sect:cond2}

We now prove that the inequality  \eqref{BDineq} is necessary for $N$-representability.
When $\gamma$ has the form \eqref{dm} and \eqref{BDeq} holds, its pre-image 
can be written in the form
\be   \label{x}
  |\Psi \ket  \!& = & \! x_{000} [\phi_1,\phi_2,\phi_3] + x_{001} [\phi_1,\phi_2,\phi_4] +
      x_{010} [\phi_1,\phi_5,\phi_3] + x_{011} [\phi_1,\phi_5,\phi_4]  \nn  \\  
      & ~ & +  ~
       x_{100} [\phi_6,\phi_2,\phi_3] + x_{101} [\phi_6,\phi_2,\phi_4] +
      x_{110} [\phi_6,\phi_5,\phi_3] + x_{111} [\phi_6,\phi_5,\phi_4].  \qquad
\ee
In this form, there is no loss of generality in assuming that the $\lambda_k$
are arranged in non-increasing order.   If we now define
\be
    S = \pmx  x_{000} & x_{001} \\  x_{010} & x_{011}  \emx \qquad 
   T = \pmx  x_{100} & x_{101} \\  x_{110} & x_{111}  \emx
\ee
then  the reduced density matrix of $ |\Psi \ket  $ is (up to a permutation)
 $W_1 \oplus W_2   \oplus W_3$ with
 \be 
     W_1 ~  = & \pmx \lambda_1 & 0 \\ 0 & \lambda_6 \emx  &  =  \pmx \tr S S^\dag & \tr S T^\dag
                                                        \\      \tr  T S^\dag & \tr TT^\dag \emx     \label{w1} \\
     W_2 ~  = &  \pmx \lambda_2 & 0 \\ 0 & \lambda_5 \emx    &  = ~~ SS^\dag + T T^\dag    \label{w2} \\
     W_3 ~ =   &  \pmx \lambda_3 & 0 \\ 0 & \lambda_4 \emx     &  =  ~~ S^\dag S +  T^\dag  T     \label{w3}                                         
 \ee  
It follows from \eqref{w1} that the eigenvalues of $SS^\dag$, 
which are the same as those of $S^\dag S$
can be written as  $\sigma, \lambda_1 \mm \sigma$ with  $0 \leq \sigma \leq \lambda_1$;
similarly those of $TT^\dag$and  $T^\dag T$
can be written as  $\tau, \lambda_6 \mm \tau$ with  $0 \leq \tau \leq \lambda_6$.

The form of \eqref{w2} and   \eqref{w3}    is suggestive of Weyl's problem  with
$A = SS^\dag,  B = TT^\dag,  C = W_2$ in the  case of  \eqref{w2} and adjoints 
reversed for  \eqref{w3}.
Weyl \cite{HJ1,W}  used the max-min principle to find necessary conditions  
\be
    a_1 + b_1 \geq c_1, \qquad a_2 + b_1 \geq c_2, \qquad  a_1 + b_2 \geq c_2 
\ee
(with all three sequences in non-increasing order).
For $2 \times 2$ matrices satisfying $\tr A + \tr B = \tr C$, these are also sufficient. 
  We apply Weyl's inequalites   to \eqref{w2} and \eqref{w3} and retain the 
stronger in each pair to obtain
\bsq \be
      \sigma + \tau  & \geq & \lambda_2 \\
      \lambda_1 \mm \sigma + \tau  & \geq  & \lambda_4  \\
      \sigma + \lambda_6 \mm \tau  & \geq  & \lambda_4 
\ee \esq 
Adding together the first two inequalities implies
\be
    2\tau \geq   \lambda_2 + \lambda_4 - \lambda_1.  
\ee
Combining this with $ 2 \lambda_6 \geq 2t  $ and using \eqref{BDeq} gives
\be
   2(1- \lambda_1) = 2 \lambda_6 \geq   \lambda_2 + 1 -  \lambda_3 - \lambda_1
\ee
which is equivalent to \eqref{BDineq}.

\section{Sufficiency} \label{sect:suff}

To prove  sufficiency, it suffices to consider a pre-image of the form
\be  \label{suffform}
  \Psi & = &    \hat{a} [\phi_1, \phi_2, \phi_3 ]+ \hat{b} [\phi_1, \phi_4, \phi_5 ] +
            \hat{s}  [\phi_6, \phi_2, \phi_4 ] + \hat{t }[\phi_6, \phi_5, \phi_3 ]
\ee
and observe that its first order reduced density matrix is diagonal in the
basis $\phi_k$ with
\bee
|\hat{a}|^2  + |\hat{b}|^2  = \lambda_1   \qquad  |\hat{s}|^2  + |\hat{t}|^2  = \lambda_6.
\eee
  Under the assumption that \eqref{BDeq} holds, the linear
  relation between the eigenvalues of $\gamma$ and 
$|\hat{a}|^2 ,  |\hat{b}|^2 ,  |\hat{s}|^2 ,  |\hat{t}|^2$   can
be inverted to yield
\bsq \label{soln} \be
   |\hat{a}|^2 =  \half  \big(\lambda_2 + \lambda_3 - \lambda_6 \big) & \quad & 
       |\hat{b}|^2 =  \half  \big(\lambda_1 - \lambda_2 + \lambda_4 \big)    \\
         |\hat{s}|^2 = \half \big(\lambda_2 - \lambda_3 + \lambda_6 \big) & \quad & 
     |\hat{t}|^2 =\half \big( \lambda_6 - \lambda_2 + \lambda_3 \big) . 
\ee   \esq
With the  ordering convention $\lambda_k \geq \lambda_{k+1}$, 
the expressions for $  |\hat{a}|^2,   |\hat{b}|^2$
and $  |\hat{s}|^2$ are all positive; and   $  |\hat{t}|^2 \geq 0 $
 is equivalent to  \eqref{BDineq}.
 
In Section~\ref{sect:cond1}, we showed slightly more
than that \eqref{BDeq} holds.   We also showed that  
  the pre-image can always be written in a form in which only 
  four of the coefficients in
\eqref{x} are non-zero.    However, neither of these forms is equivalent
to \eqref{suffform} with  $\lambda_k$ decreasing.    The equations
for the coefficients in one of those forms could have solutions only
when a stronger inequality than \eqref{BDineq} holds.    In
particular, the form obtained from \eqref{colealt} in the paragraph
before \eqref{eval24} has solutions only when
$ \lambda_1 + \lambda_2 \leq \lambda_4 + 1$.

\section{General $R = N + 3$ with $N$ odd} \label{sect:R=N+3}

It is tempting to try to extend the argument in Section~ \ref{sect:cond1} to the
general case of $ \linebreak R = N + 3$ when $N$ is odd.     Using \eqref{cole}
we can conclude as before 
that $\gamma_2$   must be $N$-representable with $R = N + 2$ and thus
has an eigenvector $|g_1\ket $ with eigenvalue $1$.  We can
write its pre-image as
\be
   |\Phi_2 \ket  & =  & a_{m} [g_1,g_2, g_3, \ldots g_{N-1}, g_{N}] + \ldots +
     a_k      [g_1,g_2, g_3, \ldots g_{2k-1} g_{2k+2} \ldots g_{N-1}, g_{N}] \qquad \nn \\
        & ~ & ~ +\ldots    +
     a_1  [g_1,g_4, g_4, \ldots g_{N+1}, g_{N+2}]          
\ee 
where $m = \half(N+1)$ and $a_k$ is the coffecient of the Slater determinant
which does {\em not} contain $g_{2k} $ or $g_{2k+1}$.    However, it
is not evident that the strong  orthogonality condition
$\bra g_1, \Phi_1 \ket_1 = 0$ holds as was the case for $N = 3$.   
 If we knew  that  
\be \label{gP}
 \lambda_1 + \bra g_1, \gamma \, g_1 \ket \leq 1,
 \ee
  strong orthogonality  would follow, and
  we could again conclude
   that $g_1$ is an eigenvector of $\gamma$ with eigenvalue 
   $\bra g_1, \gamma \, g_1 \ket  = 1 - \lambda_1$.  

      For the case $N = 5, R = N+3$ with $N$ odd, Altunbulak and Klyachko \cite{AK}  have
   shown that  $\lambda_1 + \lambda_R \leq 1$.
   This is not equivalent to  \eqref{gP}   because we
   don't know that $g_1$ is the eigenfunction for $\lambda_R$.
     A condition of the form   $\lambda_j + \lambda_{j^\prime} \leq 1$
      is sometimes called a ``strong Pauli condition''.    
  When the largest eigenvalue  is non-degenerate, we can show that
  strong orthogonality implies a strong Pauli condition with equality.
  This suggests the following 
  \begin{conj}
When $N$ is odd and $R = N + 3$,  a necessary condition for pure state
$N$-representability of a one-particle density matrix is $\lambda_1 + \lambda_R = 1$,
where we have assumed that the eigenvalues are in non-increasing order.
\end{conj}
 
     \begin{prop}
  Let $R = N + 3$ with $N$ odd and consider the decomposition \eqref{cole}
  of a one-particle density matrix $\gamma$ under the assumption that $\lambda_1$ 
  is the largest eigenvalue.    Then $|\Phi_2 \ket$ has an  eigenvector $|g_1\ket $  
  with eigenvalue $1$.   If $\bra g_1 , \Phi_1 \ket_1 = 0$, then $|g_1\ket $ is
  an eigenvector of $\gamma$ with eigenvalue $1 - \lambda_1$ and this is the
  smallest eigenvalue of $\gamma$.
   \end{prop}
   \pf   Let $|\phi_k\ket$ denote an eigenvector of $\gamma$ orthogonal to
   both $|\phi_1 \ket$ and $|g_1\ket $, and write
   \bee
       | \Phi_1 \ket & = & a \, {\cal A} | \phi_k \ot \chi_1 \ket + \sqrt{1-a^2}\,  |\psi_1 \ket  \\
        | \Phi_2 \ket & = & b\, {\cal A} |g_1  \ot  \phi_k \ot \chi_2 \ket + \sqrt{1-b^2} \, | g_1 \ot \psi_2 \ket .
   \eee
where we have absorbed any phases into $\psi_j$.
Then $\lambda_k = \lambda_1 a^2 + (1-\lambda_1)  b^2   $.   Since
each $|\psi_j \ket$ is strongly orthogonal to $|\phi_1 \ket, |g_1\ket$  and 
$|\phi_k \ket$, each
$|\psi_j \ket$ is an $(N \mm 1)$-particle function with one-rank at most $N$.
It is well-known \cite{Cole,F,RuskPH} that this implies that
$|\psi_j \ket$ is a single Slater determinant.   Since both $|\psi_1 \ket$ and
$|\psi_2 \ket$ have one-particle density matrices in the same
 $N$-dimensional subspace, it follows that the ranges
of these one-particle density matrices have a non-zero intersection.
Let $|f \ket  $ be in this intersection.   Then
\be
\bra f, \gamma  \, f \ket = (1-a^2) \lambda_1 + (1-b^2) (1- \lambda_1) = 1 - \lambda_k
\ee
Thus, if, $\lambda_k < 1- \lambda_1 $, then $\bra f, \gamma  \, f \ket  > \lambda_1$
contradicting the assumption that $\lambda_1$ is the largest eigenvalue.  \qed
 
 \section{Further connections with Weyl's problem}   \label{sect:Weyl}

Now assume that  $g_1$ is strongly orthogonal to $\Phi_1$ and, as
 in  \eqref{colealt}, write 
\be    \label{colealt2}
    \gamma = \lambda_1 \proj{\phi_1} + (1- \lambda_1) \proj{g_1} +
             \lambda_1 \gamma_1 + (1 -  \lambda_1) \wtd{\gamma}_2.
\ee
The $N$-representability problem in this situation
 is reduced to finding conditions which ensure that a density
 matrix is a convex combination of two $(N \mm 1)$-representable density
 matrices of rank $N+1$ which satisfy  an additional orthogonality constraint.
 Write 
 \bee
     | \Phi_1 \ket & = &  \sum_{k_1 < k_2 < \ldots k_{N-1}}  x_{k_1  k_2  \ldots k_{N-1}}
         [g_1,g_2, \ldots g_{N-1}]  \\
          | \Phi_2 \ket & = &\sum_{k_1 < k_2 < \ldots k_{N-1}}  y_{k_1  k_2  \ldots k_{N-1}}.
         [g_1,g_2, \ldots g_{N-1}] 
 \eee
Let $X,Y$ be the corresponding anti-symmetric tensors, and let
 \be
 XY^\dag =  \! \! \sum_{k_2, k_3 \ldots k_M}   \!
     x_{k_1,k_2, \ldots k_M} \, \overline{y}_{k_1,k_2, \ldots k_M}
     \ee 
     denote contraction over    $k_2 \ldots k_M$.   Then, we can rewrite \eqref{colealt}
as      \be  \label{constr}
     \gamma - \lambda_1 \proj{\phi_1} - (1- \lambda_1) \proj{g_1} = XX^\dag + YY^\dag
  \ee 
 with the constraint  $\bra \Phi_1 , \Phi_2 \ket = \tr XY^\dag = 0$.
This is a constrained version of Weyl's problem.
 If the $R = N+3$ problem could be solved in this way,
  then by particle-hole duality, we would also have the solution to the
  $3$-representability problem.   
   Although we we do not know if strongly orthogonality of $g_1$    to $\Phi_1$
   holds in general, this viewpoint provides a connection to Weyl's problem 
   that is more general the situation for which it was used  in Section~\ref{sect:cond2}.

 For general $R$ (or for $R = N+3$  without the simplification that leads 
to \eqref{colealt}),
 Coleman's  Lemma~\ref{thm:cole} gives a constrained version of
 Weyl's problem with  $\gamma_1 = XX^\dag$ and $\gamma_2 = YY^\dag$.
 But now $\gamma_1$ is $(N \mm 1)$-representable and $\gamma_2$
 is $N$-representable and the orthogonality condition \eqref{stong} must
 be translated to tensors of different size.    Nevertheless, it now seems clear that
 what Coleman referred to as a double induction lemma, was a constrained
 version of Weyl's problem.    The solution to Weyl's problem was given
 less than 10 years ago, with more recent refinements \cite{KT}.  Thus,
  it is not  surprising that the pure state $N$-representability problem also 
  resisted solution  and that Klyachko succeeded by using powerful techniques
  associated with Schubert calculus to solve both problems.
  
  \bigskip
   
\noindent  {\bf Acknowledgment:}    It is a pleasure to recall that most of this
work was the result of discussions with R.E. Borland and K. Dennis during
a visit to the National Physical Laboratory in Great Britain in the fall of 1970.
 
 \pagebreak

 \end{document}